\documentclass[reprint,showpacs]{revtex4-1}
\usepackage{amssymb,amsmath}
\usepackage{graphicx}

\usepackage{bm}
\makeatother
\begin{document}

\title{Investigating Isomorphs with the Topological Cluster Classification}
\author{Alex Malins}
\affiliation{Bristol Centre for Complexity Sciences, University of Bristol, Bristol, BS8 1TS, UK}
\affiliation{School of Chemistry, University of Bristol, Cantock's Close, Bristol, BS8 1TS, UK}
\author{Jens Eggers}
\affiliation{School of Mathematics, University of Bristol, University Walk, Bristol, BS8 1TW, UK}
\author{C. Patrick Royall}
\affiliation{HH Wills Physics Laboratory, Tyndall Avenue, Bristol, BS8 1TL, UK}
\affiliation{School of Chemistry, University of Bristol, Cantock's Close, Bristol, BS8 1TS, UK}
\affiliation{Centre for Nanoscience and Quantum Information, Tyndall Avenue, Bristol, BS8 1FD, UK}

\email{paddy.royall@bristol.ac.uk}

\date{\today}

\begin{abstract}
Isomorphs are lines in the density-temperature plane of certain ``strongly-correlating'' or ``Roskilde simple'' liquids where two-point structure and dynamics have been shown to be close to identical up to a scale transformation. Here we consider such a liquid, a Lennard-Jones glassformer, and investigate the behavior along isomorphs of higher-order structural and dynamical correlations. We then consider an inverse power law reference system mapped to the Lennard-Jones system [Pedersen \textit{et al.}, Phys.\ Rev.\ Lett.\ \textbf{105}, 157801 (2010)]. Using the topological cluster classification to identify higher-order structures, in both systems we find bicapped square anti-prisms, which are known to be a locally favored structure in the Lennard-Jones glassformer. The population of these locally favored structures is up to 80\% higher in the Lennard-Jones system than the equivalent inverse power law system. The structural relaxation time of the two systems, on the other hand, is almost identical, and the four-point dynamical susceptibility is marginally higher in the inverse power law system. Upon cooling the lifetime of the locally favored structures in the Lennard-Jones system are up to 40\% higher relative to the reference system.
\end{abstract}

\maketitle

\section{Introduction}

The claim that many of the structural properties of liquids derive from repulsive hard core-like interactions, which often are of an inverse power law form, is a concept which has a considerable history~\cite{widom1967,weeks1971,chandler1983,barker1976}. Thus one might argue that by removing the attractive interactions, thereby arriving at an inverse power law (IPL) model, any structural properties, including those related to glassy behavior at low temperatures, are preserved. The advantage is that IPL
models exhibit scale invariance, which maps points in the phase diagram to equivalent points along so-called isomorph lines~\cite{hansen,branka2006,gnan2009}. This results in a considerable conceptual simplification of the low-temperature behavior and, in particular, reduces the phase diagram to a 1-dimensional representation. In addition to structural and thermodynamic quantities, dynamical quantities such as the relaxation time can also be mapped from one system to another for both Newtonian dynamics~\cite{young2003,young2005} and Brownian dynamics~\cite{guevara2003,schmiedeberg2011}. Furthermore the two classes of dynamics can be mapped to one another~\cite{pond2011,lopez2012}.

Recently this concept of scaling has been taken a step further and applied to models such as a Lennard-Jones (LJ) glassformer, for which no exact mapping exists. Lines previously related by scale invariance are replaced by isomorphs in the phase diagram, along which any two points are {\it approximately} equivalent. A convenient way to measure the degree to which this invariance holds is to observe to what degree the fluctuations of the potential energy $U$ and the virial $W$ (see Eq.~\ref{eqPV}) in constant volume 
ensembles are correlated~\cite{bailey2008a,bailey2008,schroder2009,gnan2009,schroder2011}. While IPL liquids display perfect correlation between the fluctuations of $U$ and $W$, it is proposed in these papers that liquids which exhibit a high degree of correlation obey a number of scaling laws for their static structural, dynamical and thermodynamic properties. Such liquids have been termed ``strongly correlating'' or ``Roskilde simple'' liquids~\cite{ingebrigtsen2012,separdar2013}. A perspective on this approach has recently appeared~\cite{dyre2013}, and a paper summary of the theory is available in reference ~\cite{dyre2013a}.        

A connection is made between Roskilde simple liquids with attractions and purely repulsive IPL liquids by the gradient $\gamma$ of a linear fit for a scatter plot of $U$ and $W$ fluctuations. An IPL reference potential with exponent $n=3\gamma$ reproduces to a good approximation the pair structure and dynamics for the full LJ system with attractions at the state point where $\gamma$ is measured~\cite{gnan2009,pedersen2011}. Moreover for systems with good isomorphs through the phase diagram, it has been conjectured that $\gamma$ is a density-dependent parameter, i.e.\ $\gamma=\gamma(\rho)$, and as such the IPL approximation remains valid along the isochor. This conjecture has been shown to break down however for liquids where the potential is truncated at ranges shorter than the mean inter-particle separation~\cite{berthier2011f,ingebrigtsen2012}, or if the correlation between $U$ and $W$ drops significantly along the isochor such that the liquid is no longer strongly correlating.

Pedersen \textit{et al.\ }employed this conjecture to derive a repulsive IPL potential that approximates the attractive Kob-Andersen (KA) Lennard-Jones glass former~\cite{pedersen2010c}. The IPL system matched the pair structure, dynamics, and fragility of the KA mixture along the $\rho=1.2$ isochor. A system with interactions truncated at the minimum of the potential following Weeks, Chandler and Andersen (WCA)~\cite{weeks1971} did not reproduce the dynamics, although its pair structure was very similar~\cite{berthier2011}.
The correspondence between the purely repulsive IPL system and the Lennard-Jones system with attractions held its validity over a range of temperatures from within the Arrhenius liquid regime to deeply supercooled states. 

In order to assess the usefulness of these ideas as applied to a phenomenon such as the slowing down of a glass former upon cooling, it is crucial to measure how well the proposed equivalence applies to quantities of possible physical relevance, such as those measuring the structural ordering of liquids~\cite{malins2013jcp,malins2013fara}. Thus far numerical investigations into the invariance of structure and dynamics along isomorphs focused on two-point correlations, i.e.\ $g(r)$ 
and $F_\mathrm{s}(k,t)$~\cite{gnan2009,pedersen2010c}. In this paper we go one stage further and test the validity of isomorphs using higher order correlation functions for structure and dynamics measured with the topological cluster classification~\cite{williams2007,malins2013tcc}. In particular we examine the behavior of ``11A'' bicapped square antiprism clusters of 11 particles, which have been identified as being prevalent structures in the KA supercooled liquid~\cite{coslovich2007,coslovich2011,malins2013fara}, to be slowly relaxing~\cite{malins2013fara} and related to a structural-dynamical phase transition in the same system~\cite{speck2012}. Moreover ultra-stable KA glasses formed by (\emph{in-silico}) chemical vapor deposition have been shown to be rich in 11A~\cite{singh2013}. Finally the presence and size of domains of 11A clusters has been suggested as a way to rationalize the difference in fragility between the full KA LJ system with attractions and its WCA counterpart with purely repulsive interactions~\cite{coslovich2011}.

Here we use the yield of 11A as an order parameter for structure, and the lifetime autocorrelation function as an order parameter for the many-particle dynamical correlations on cooling between the Kob-Andersen Lennard-Jones model and its mapped inverse power law reference system. We find that, although the model system was previously considered to reproduce the structure and dynamics of the full system with attractions, the 11A are in fact 80\% more numerous and exhibit a substantially longer lifetime at low temperature state points.
 
This paper is organized as follows. We briefly review key properties in strongly correlating liquids in section \ref{sectionStronglyCorrelatingLiquids}, and the means to identify isomorphs in section \ref{sectionIsomorphs}. We describe our simulations in section \ref{sectionSimulation} and our structural analysis in section \ref{sectionTCC}. In our results, we first consider the behavior of higher-order structure and dynamics along isomorphs for strongly correlating liquids in Section \ref{sectionStructure}. We then turn to the structural and dynamical analysis of the Lennard-Jones and its model inverse power law system as dynamical arrest is approached in section \ref{sectionGlassCompare}. Sections \ref{sectionDiscussion} and \ref{sectionConclusion} concern the discussion and conclusion respectively.

\section{Methods}
\label{Methods}

\subsection{Roskilde simple liquids}
\label{sectionStronglyCorrelatingLiquids}

\begin{figure*}
\begin{centering}
\includegraphics[width=18cm]{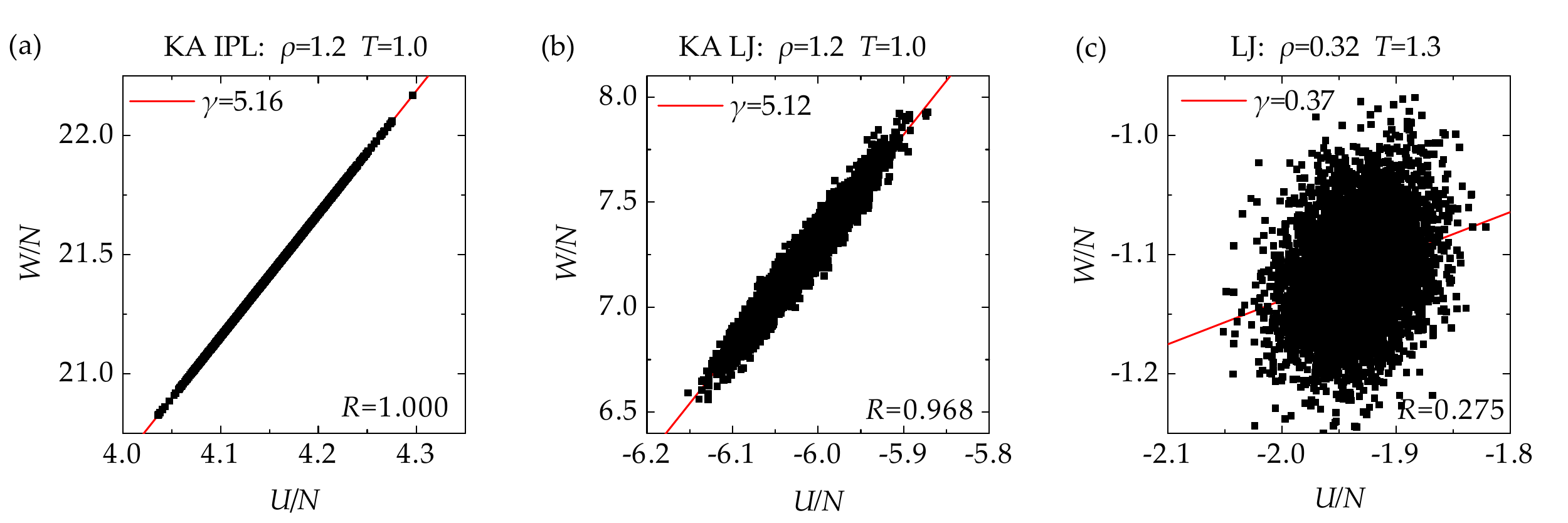}
\par\end{centering}
 \caption[Scatters plot of the potential and the virial.]{Scatters plot of the potential $U$ and the virial $W$, squares, and linear fits, red lines. (a) KA IPL $\rho=1.2$, $T=1.0$, perfect correlation of $U$ and $W$. (b) KA LJ $\rho=1.2$, $T=1.0$, strongly correlating. (c) Monodisperse LJ above the critical point, $\rho=0.32$, $T=1.3$, weakly correlating. 
\label{figWUscatter}}
\end{figure*}

In the following sections \ref{sectionStronglyCorrelatingLiquids} to \ref{sectionIsomorphs} we summarize the theory of Roskilde simple liquids and isomorphs. In the canonical $NVT$-ensemble (constant number of particles $N$, volume $V$ and temperature $T$), the pressure of a liquid $P$ is the sum of contributions from ideal gas and interaction terms:
\begin{equation}
PV=Nk_\mathrm{B}T+W,
\label{eqPV}
\end{equation}
\noindent The contributions from interactions are thus expressed in the virial term $W$ which is given by
\begin{equation}
W(\mathbf{r}^N)=-\frac{1}{3}\sum_{i=1}^N \sum_{i<j}^N r_{ij}u'(r_{ij}).
\label{eqVirial}
\end{equation}
The fluctuations of $U$ and $W$ between configurations in the $NVT$-ensemble can be represented on a scatter plot, as shown in Fig.~\ref{figWUscatter}. The degree of correlation between the variables is characterized Pearson's $R$ coefficient
\begin{equation}
R=\frac{\langle \Delta W \Delta U \rangle}{\sqrt{\langle ( \Delta W)^2\rangle} \sqrt{\langle ( \Delta U)^2\rangle}},
\label{CH7Eq:R}
\end{equation}
\noindent where $\Delta U=U-\langle U \rangle$ and $\Delta W=W-\langle W \rangle$. Roskilde simple liquids are defined empirically as liquids where $R>0.9$~\cite{bailey2008a}. The parameter that defines the ``slope'' of the correlation between the $U$ and $W$ fluctuations is $\gamma$
\begin{equation}
\gamma=\frac{\langle \Delta W\Delta U\rangle}{\langle ( \Delta U)^2\rangle}.
\label{CH7Eq:gamma}
\end{equation}
The parameters $R$ and $\gamma$ are state point dependent and liquids may demonstrate the strongly correlating property ($R>0.9$) in some regions of phase space and not in others. For example, in the vicinity of the critical point, liquids are not strongly correlating~\cite{dyre2013}.
In Figs.~\ref{figWUscatter}(a) to (c) we show scatter plots of $W/N$ versus $U/N$ for an IPL system, a strongly correlating liquid state point (KA LJ) and a weakly correlating liquid state point (monodisperse LJ) respectively. Perfect correlation is obtained for the IPL liquid as $W=\gamma U$ strictly for IPL potentials [Fig.~\ref{figWUscatter}(a)].

\subsection{Isomorphs}
\label{sectionIsomorphs}

Isomorphs are curves through regions of the phase diagram where there are strong correlations between the fluctuations of $U$ and $W$. A number of structural, dynamical and thermodynamic quantities are invariant at state points along isomorphs when presented in a set of reduced units. The theory can be reduced to a single approximation for the relative Boltzmann weights that identical microstates of two ``isomorphic'' state points \footnote{\noindent Isomorphic state points are found on the same isomorph.} contribute to the partition function.  Two state points $(\rho_1,T_1)$ and $(\rho_2,T_2)$ are isomorphic if the statistical weights of representative microscopic configurations at state point $1$ ($\mathbf{r}^N_{(1)}$) are proportional to the statistical weights of scaled configurations of state point $1$ at state point $2$, where the configurations are scaled between the state points as $\mathbf{r}^N_{(2)}=\rho_1^{1/3}\mathbf{r}^N_{(1)}/\rho_2^{1/3}$, i.e.~\citep{gnan2009}:
\begin{equation} 
\exp\left[-U(\mathbf{r}_{(1)}^N)/ k_{\mathrm{B}} T_{1}\right] = C_{12}\exp\left[-U(\mathbf{r}_{(2)}^N)/k_{\mathrm{B}} T_{2}\right].
\label{CH7Eq:Boltzmann_factor_IPL}
\end{equation}
\noindent The parameter $C_{12}$ only depends on the state points 1 and 2 and not on the details of the microscopic configurations.
The theory of isomorphs is rigorous when $C_{12}=1$, which only occurs for IPL potentials. For other potentials the isomorphs are approximate, where the strength of the approximation depends on how close $R$ is to unity between state points on an isomorphic curve.

The consequence of Eq.~\ref{CH7Eq:Boltzmann_factor_IPL} is that a number of quantities measured along isomorphs can be collapsed when cast in terms of reduced units for energy ($k_\mathrm{B}T$), length ($\rho^{-1/3}$) and time ($\sqrt{m/k_\mathrm{B}T\rho^{2/3}}$ for Newtonian dynamics). As the microstates of isomorphic state points scale trivially onto each other, all structural measurements are invariant to a good approximation. Newtonian dynamics in reduced time are invariant between isomorphic state points as long as the timescale of the thermostat is adjusted to be invariant between the state points. Invariant thermodynamic properties along isomorphs include the excess entropy, the configurational entropy and the isochoric specific heat~\cite{gnan2009}.

The shape of isomorphic curves can be calculated for potentials that are 
the sum of IPLs with differing prefactors $A_\mathrm{IPL}$ and exponents $\gamma$ using the method outlined in reference~\cite{bohling2012}. For LJ potentials, as employed by the KA model, a scaling function is defined relative to a reference state point for the isomorph with $(\rho^\ast,T^\ast)$ and $\gamma^\ast$. The scaling function $h(\tilde{\rho})$ is
\begin{equation} 
h(\tilde{\rho})=\tilde{\rho}^4(\gamma^\ast/2-1)-\tilde{\rho}^2(\gamma^\ast/2-2),
\label{CH7Eq:scaling_function}
\end{equation}
\noindent where $\tilde{\rho}=\rho/\rho^\ast$. $h(\tilde{\rho})/T$ is constant along isomorphs meaning that all state points along an isomorph can then be calculated from the value of $h(\tilde{\rho})/T$ at the reference state point, i.e.\ $h(\tilde{\rho})/T=h(1)/T^\ast$.

\subsection{Simulation details}
\label{sectionSimulation}

We consider the isomorphs of the KA LJ mixture and its reference KA IPL system defined in reference~\cite{pedersen2010c} in 3D molecular dynamics simulations with periodic boundary conditions. The IPL potential reads
\begin{equation}
u_{\alpha\beta}(r)=A_\mathrm{IPL} \varepsilon_{\alpha\beta} \left( \frac{\sigma_{\alpha\beta}}{r} \right) ^{3\gamma},
\label{CH7Eq:u_IPL}
\end{equation}
\noindent where the parameters $A_\mathrm{IPL}=1.9341$ and $\gamma=5.16$ are calculated from our simulations. The values of these parameters are calculated respectively from gradients of scatter plots of the LJ virial versus the LJ potential, and the LJ potential versus IPL potential without the $A_\mathrm{IPL}$ prefactor, across the state points $\rho^\ast=1.2$ and $T^\ast$ values of 0.5, 0.525, 0.6, 0.75, 1.0 and 2.0. The subscripts $\alpha$ and $\beta$ refer to the different species in the KA binary mixture which for both IPL and LJ systems is composed of 80\% large ($A$) and 20\% small ($B$) species particles of the same mass $m$~\cite{kob1995a}. The nonadditive interactions between each species, and the cross interaction, are given by $\sigma_{AA}=\sigma$, $\sigma_{AB}=0.8\sigma$, $\sigma_{BB}=0.88\sigma$, $\epsilon_{AA}=\epsilon$, $\epsilon_{AB}=1.5\epsilon$, and $\epsilon_{BB}=0.5\epsilon$. The results are quoted in reduced units with respect to the $A$ particles, i.e.\ we measure length in units of $\sigma$, energy in units of $\epsilon$, time in units of $\sqrt{m\sigma^2/\epsilon}$, and set Boltzmann's constant $k_\mathrm{B}$ to unity. 

We truncate and shift both LJ and IPL potentials at $2.5\sigma_{\alpha\beta}$ and simulate in the $NVT$-ensemble for a system of $N=1000$ particles, $N_A=800$. The temperature is controlled using the Nos\'{e}-Hoover thermostat with coupling constant $\tau_{\mathrm{NH}}=0.03$. The values of $\tau_{\mathrm{NH}}$ are adjusted along the isomorphs such that they are identical when reduced by $\sqrt{m/k_\mathrm{B}T\rho^{2/3}}$. The initial configuration at $\rho=1.2$, $T=2.0$ is a face-centred cubic crystal which is melted and the liquid equilibrated before trajectories are sampled. Each lower temperature state point on the $\rho=1.2$ isochor is thoroughly equilibrated after an instantaneous quench from the previous higher $T$ state point. Six trajectories of $300\tau_\alpha^A$ length are sampled for each state point, where $\tau_\alpha^A$ is the alpha-relaxation time of the large particles. We determine $\tau_\alpha^A$ by fitting a stretched exponential (Kohlrausch-Williams-Watts) form to intermediate scattering functions (ISFs) of the $A$-species such as those in Fig.~\ref{figGdyn}(a). The trajectories consist of $3000$ configurations, where each configuration is separated by $\tau_\alpha^A/10$.

We consider six isomorphs for both the KA LJ and KA IPL models, where the reference state points are $\rho^\ast=1.2$ with $T^\ast=$\ 0.5, 0.525, 0.6, 0.75, 1.0 and 2.0. The initial configuration for a new state point along an isomorph is obtained by scaling the positions and velocities from a single configuration at the reference state point, using $\mathbf{r}_i=(\rho^\ast)^{1/3}\mathbf{r}^\ast_i/\rho^{1/3}$ and $\mathbf{v}_i=T^{1/2}\mathbf{v}^\ast_i/(T^\ast)^{1/2}$. The new state point is equilibrated for at least $300\tau_\alpha^A$ before sampling six trajectories, each $300\tau_\alpha^A$ in length, for analysis. The long equilibration time is sufficient to ensure that there is no ageing of the system once the trajectories are sampled. This is checked by ensuring there is no evolution in the pair correlation functions, ISFs, and number of clusters identified by the topological cluster classification during the trajectories. The equilibration of the isomorphic state points on the $\rho=2.0$ isochor for the KA LJ system is checked independently by equilibrating a liquid at high $T$ for $\rho=2.0$, and then quenching down slowly to the lower temperatures. The results obtained with the quenching protocol match those measured for the state points obtained by scaling along the isomorphs from $\rho^\ast=1.2$.

Another benefit of the equilibration is that it ensures the trajectories sampled are statistically independent from the initial configuration of the $\rho^\ast=1.2$ reference state point, and from the trajectories at other isomorphic state points \footnote{\noindent The dynamics for isomorphic state points of the IPL system are in theory invariant in reduced units, as they originate from the same initial configuration at the reference state point. However the build-up of computational round-off errors during the initial equilibration period before trajectories are sampled, and round-off errors due to the scaling of the positions and the velocities from the initial configuration to the isomorphic state points, ensure that the trajectories sampled at different isomorphic state points are statistically independent.}. We traverse out isomorphs from $\rho^\ast=1.2$ to $\rho=$\ 1.0, 1.4, 1.6, 1.8, and 2.0. The $\rho=1.0$ state points for the KA LJ model crystallize on all isomorphs except $\rho^\ast=1.2$, $T^\ast=2.0$, indicating that the crystallization time along isomorphs is not invariant.

When comparing higher-order structure and dynamics between the KA LJ and KA IPL models on the $\rho=1.2$ isochor, we consider temperatures $T=$\ 0.45, 0.5, 0.525, 0.6, 0.75, 1.0 and 2.0. Due to the long runtime of the simulations at the $T=0.45$ state points, only two trajectories are sampled for analysis after the quench and initial equilibration period ($>300\tau_\alpha^A$).

\begin{figure*}
\begin{centering}
\includegraphics[width=18cm]{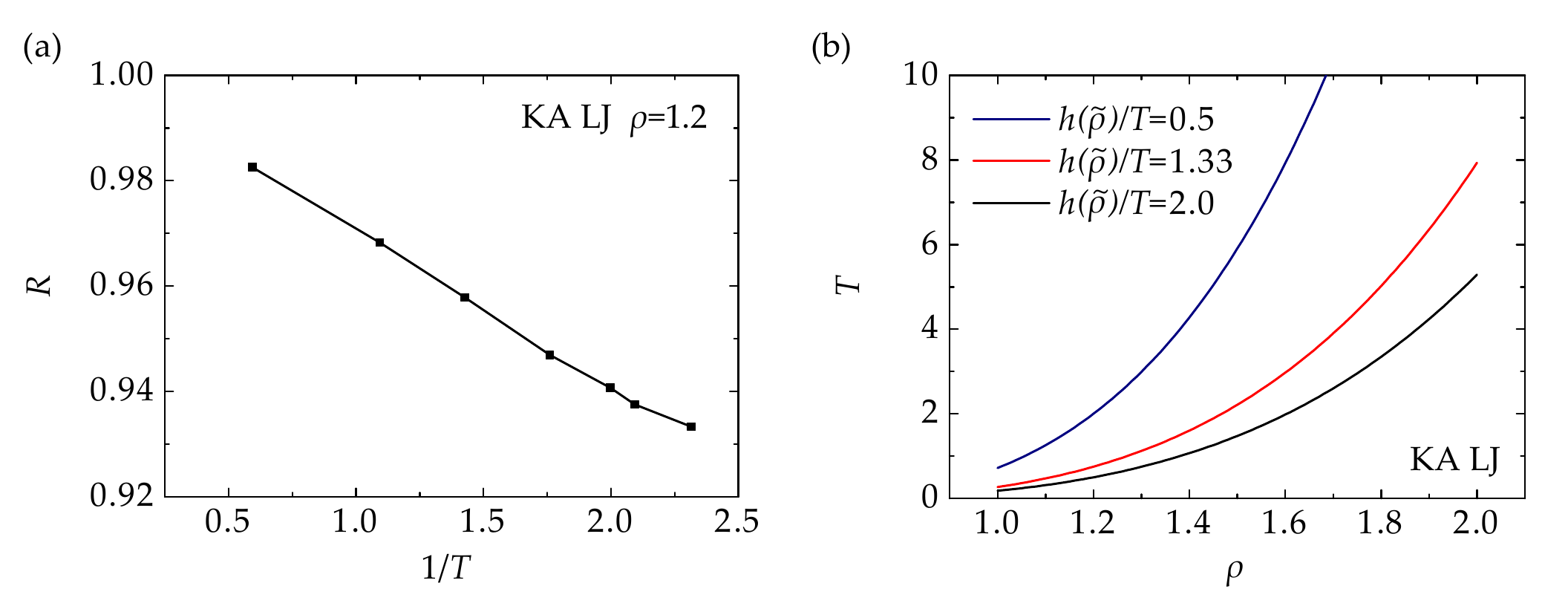}
\par\end{centering}
 \caption{Isomorphs in the KA LJ phase diagram. (a) The correlation coefficient for $U$-$W$ fluctuations along the $\rho=1.2$ isochor. (b) The shape of isomorphs in the phase diagram are described by $h(\tilde{\rho})\propto T$.
\label{figGRIsomorphs}}
\end{figure*}

In Fig.~\ref{figGRIsomorphs}(a) we show the correlation coefficient $R$ for the $U$-$W$ fluctuations in the KA LJ mixture. As the temperature decreases the value of $R$ decreases in an approximately linear fashion in $1/T$ for our parameters.
In Fig.~\ref{figGRIsomorphs}(b) the shape of the isomorphic curves in the KA LJ phase diagram are shown, as calculated using Eq.~\ref{CH7Eq:scaling_function}.

\subsection{Structural analysis}
\label{sectionTCC}

We analyse the way in which the particles in the supercooled liquids are structured using the topological cluster classification (TCC) algorithm~\cite{williams2007,malins2013tcc}. This algorithm identifies a number of local structures. Using the TCC, for the KA mixture we have shown that the relevant structure which forms long-lived and slow domains of particles is the bicapped square antiprism, 11A~\cite{malins2013fara}. These are predominantly formed with a small B-type particle at the centre surrounded by mostly 10, sometimes 9 and occasionally 8 larger A-type particles. This structure is also the minimum energy cluster for 11 KA particles~\cite{malins2013fara}. An 11A is illustrated in Fig.~\ref{figGIPLisomorph}(a).

The first stage of the TCC algorithm is to identify the bonds between neighboring particles. The bonds are detected using a modified Voronoi method with a maximum bond length cut-off of $r_\mathrm{c}=2.0$ for all types of interaction ($AA$, $AB$ and $BB$). A parameter which controls identification of four- as opposed to three-membered rings, $f_\mathrm{c}$, is set to unity thus yielding the direct neighbors of the standard Voronoi method~\cite{brostow1978}. Groups of 11 particles which are topologically equivalent to 11A are detected based on the bond network.
Further details can be found in~\cite{williams2007,malins2013tcc}.

The maximum bond length $r_\mathrm{c}$ in the cluster detection is greater than the longest bond length detected at any of the state points for the KA LJ and KA IPL models studied in this paper (including along the isomorphs). This means that the detection of the clusters is, in effect, independent of any length scale. Thus the configurations along isomorphs do not need to be rescaled back to the reference density $\rho^\ast=1.2$, or the parameter $r_\mathrm{c}$ scaled along the isomorphs, in order that the detection of clusters is fair and consistent between isomorphic state points.

In order to investigate the cluster lifetimes, we employ the \emph{dynamic} topological cluster classification algorithm~\cite{malins2013jcp,malins2013fara}. A lifetime $\tau_{\ell}$ is assigned to each ``instance'' of an 11A, where an instance is defined by the unique indices of the particles within the cluster. Each instance of a cluster occurs between two frames in the trajectory and the lifetime is the time difference between these frames. Any periods where the instance is not detected by the TCC algorithm are shorter than $\tau_\alpha^A$ in length, and no subset of the particles becomes un-bonded from the others during the lifetime of the instance.

\section{Results}
\subsection{Structure and dynamics across isomorphic state points}
\label{sectionStructure}

\begin{figure*}
\begin{centering}
\includegraphics[width=18 cm]{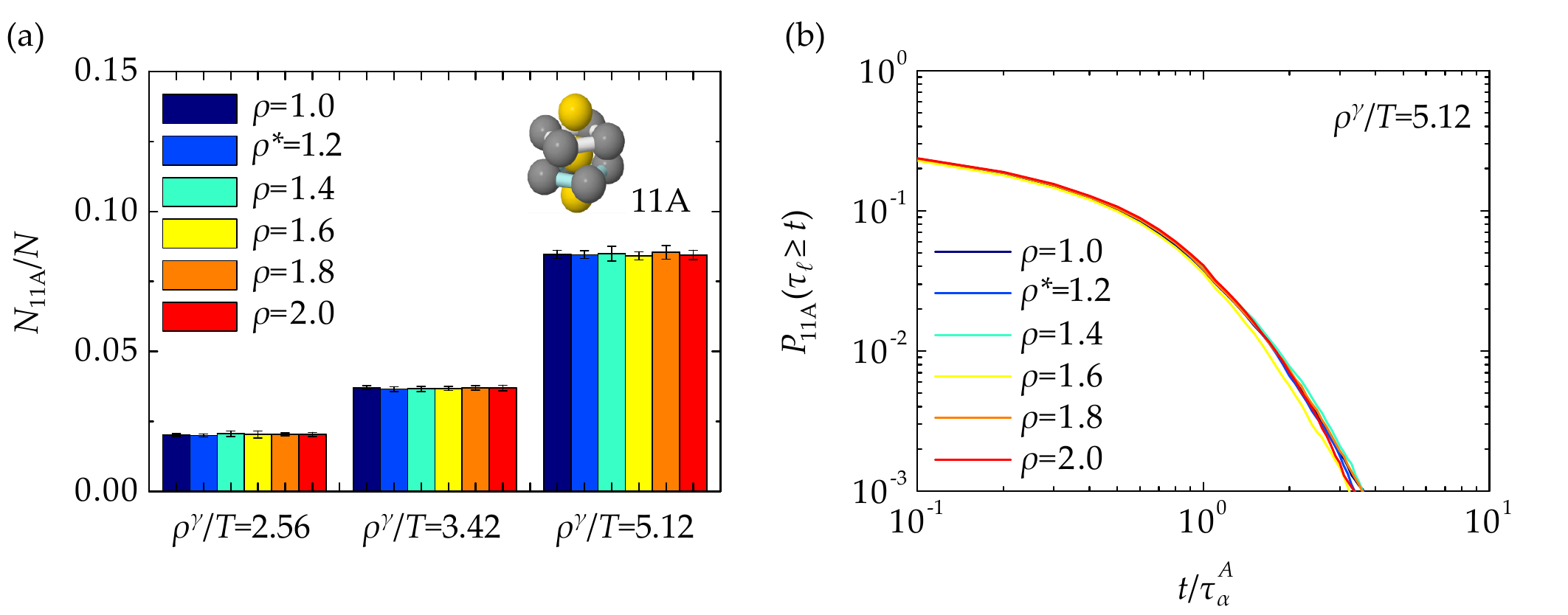}
\par\end{centering}
 \caption{The statics and dynamics of 11A clusters for the KA IPL isomorphs. As this is an IPL system the isomorphs are rigorous. (a) The fraction of particles $N_\mathrm{11A}/N$ detected within 11A clusters by the TCC algorithm. The temperatures for the reference density $\rho^\ast=1.2$ are $T=\{1.0,0.75,0.5\}$. Error bars show plus and minus two standard deviations from the mean. Inset illustrates the 11A bicapped square antiprism.
(b) The lifetime autocorrelation function $P_\mathrm{11A}(\tau_\ell\ge t)$ of the 11A clusters for the most deeply supercooled state points, $\rho^\gamma/T=5.12$.
\label{figGIPLisomorph}}
\end{figure*}

We begin by analysing the structure and dynamics along the isomorphs for the KA IPL system. In Fig.~\ref{figGIPLisomorph}(a) the number of particles detected within 11A clusters along three different isomorphs for the KA IPL system is shown. The isomorphs for this system are exact and hence $N_\mathrm{11A}/N$ is identical across all state points along an isomorph, within the statistical limits of the data. The lifetime autocorrelation functions for the 11A cluster $P_\mathrm{11A}(\tau_\ell\ge t)$ all collapse once the data have been normalized by $\tau_\alpha^A$. These results demonstrate that higher order correlations in structure and dynamics are invariant along isomorphs in IPL systems, as expected.

\begin{figure*}
\begin{centering}
\includegraphics[width=18 cm]{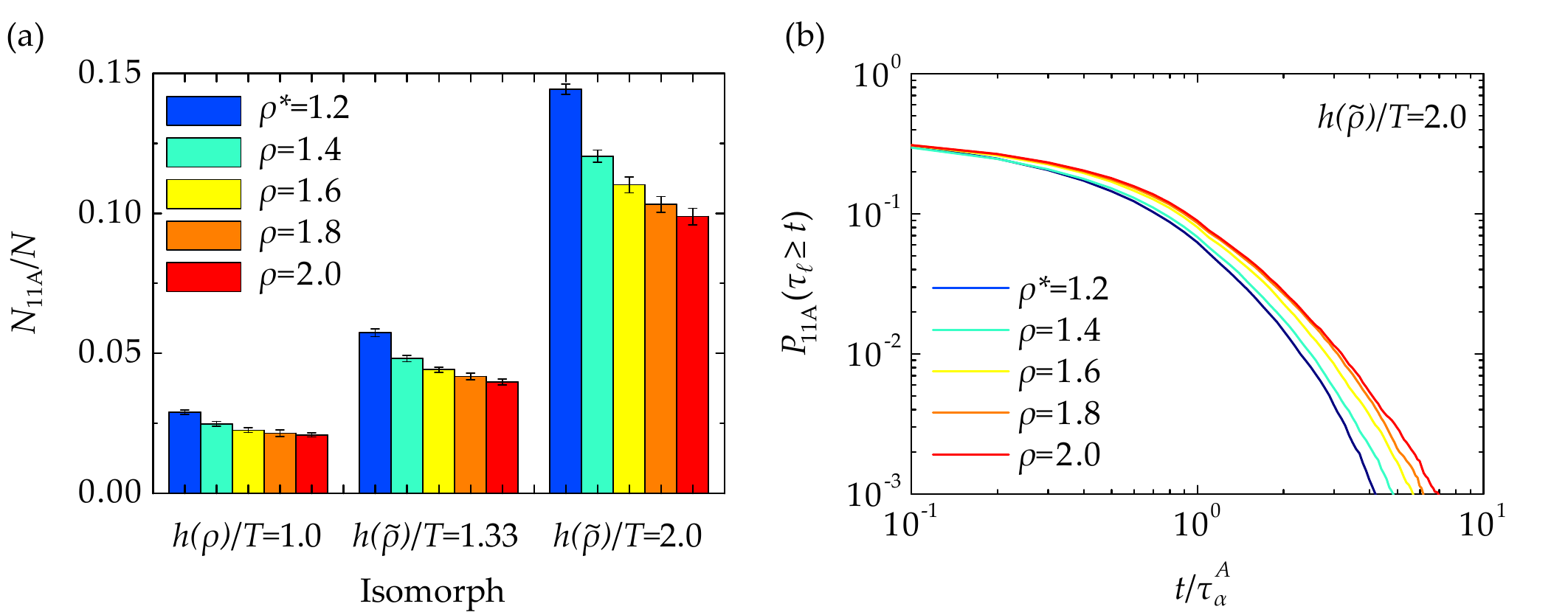}
\par\end{centering}
 \caption{The statics and dynamics of 11A clusters for the KA LJ isomorphs. (a) The fraction of particles $N_\mathrm{11A}/N$ detected within 11A clusters by the TCC algorithm. The temperatures for the reference density $\rho^\ast=1.2$ are $T=\{1.0,0.75,0.5\}$. Error bars as per Fig.~\ref{figGIPLisomorph}(a). (b) The lifetime autocorrelation function $P_\mathrm{11A}(\tau_\ell\ge t)$ of the 11A clusters for the most deeply supercooled state points, $h(\tilde{\rho})/T=2.0$.
\label{figGKAisomorph}}
\end{figure*}

We move on to the full KA LJ system where structure and dynamics are expected to be invariant along isomorphs to a good approximation. As Figs.~\ref{figGKAisomorph}(a) and (b) show, there are in fact 
\emph{continuous} changes in $N_\mathrm{11A}/N$ and $P_\mathrm{11A}(\tau_\ell\ge t)$ between isomorphic state points, yielding a difference of up to 40\% for the most deeply supercooled state points.
The relative differences in these quantities between isomorphic state points is much larger than the relative differences in two-body static and dynamic correlation functions $g(r)$ and $F_\mathrm{s}^A(k,t)$, 
cf.~reference~\cite{gnan2009}. This shows that isomorphs are perhaps less accurate in the regime of glassy dynamics, which is dominated by strong many-particle correlations.

\subsection{Comparison of the purely repulsive system to the full Lennard-Jones system}
\label{sectionGlassCompare}

\begin{figure*}
\begin{centering}
\includegraphics[width=18 cm]{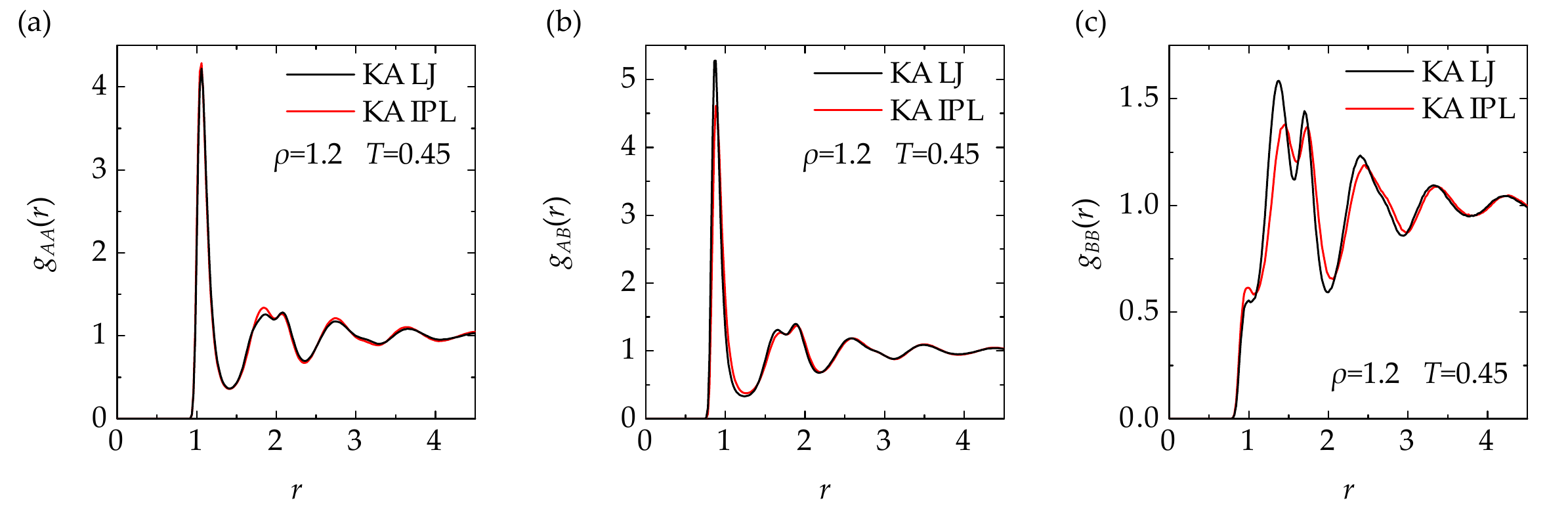}
\par\end{centering}
 \caption{Partial radial distribution functions for the KA LJ model (black lines) and its IPL reference system (red lines). The state point is $\rho=1.2$, $T=0.45$.} (a) $g_{AA}(r)$, (b) $g_{AB}(r)$, and (c) $g_{BB}(r)$. \label{figG}
\end{figure*}

In this section we compare structural and dynamical features of the KA LJ system and its purely repulsive IPL reference system. In Figs.~\ref{figG}(a) to (c) we compare the partial radial distribution functions for the systems with and without attractions. The partial radial distribution functions are broadly similar between the two models. The largest differences are seen for $g_{BB}(r)$, and around the first of peak of $g_{AB}(r)$. We note that the partial radial distribution functions from both the IPL and WCA models are good matches for those pertinent to the full LJ system~\cite{pedersen2010c}. Note that differences in $g_{BB}$ have also been observed between dynamically ``active'' and ``inactive'' trajectories in the KA mixture~\cite{speck2012jcp}.

\begin{figure*}
\begin{centering}
\includegraphics[width=18 cm]{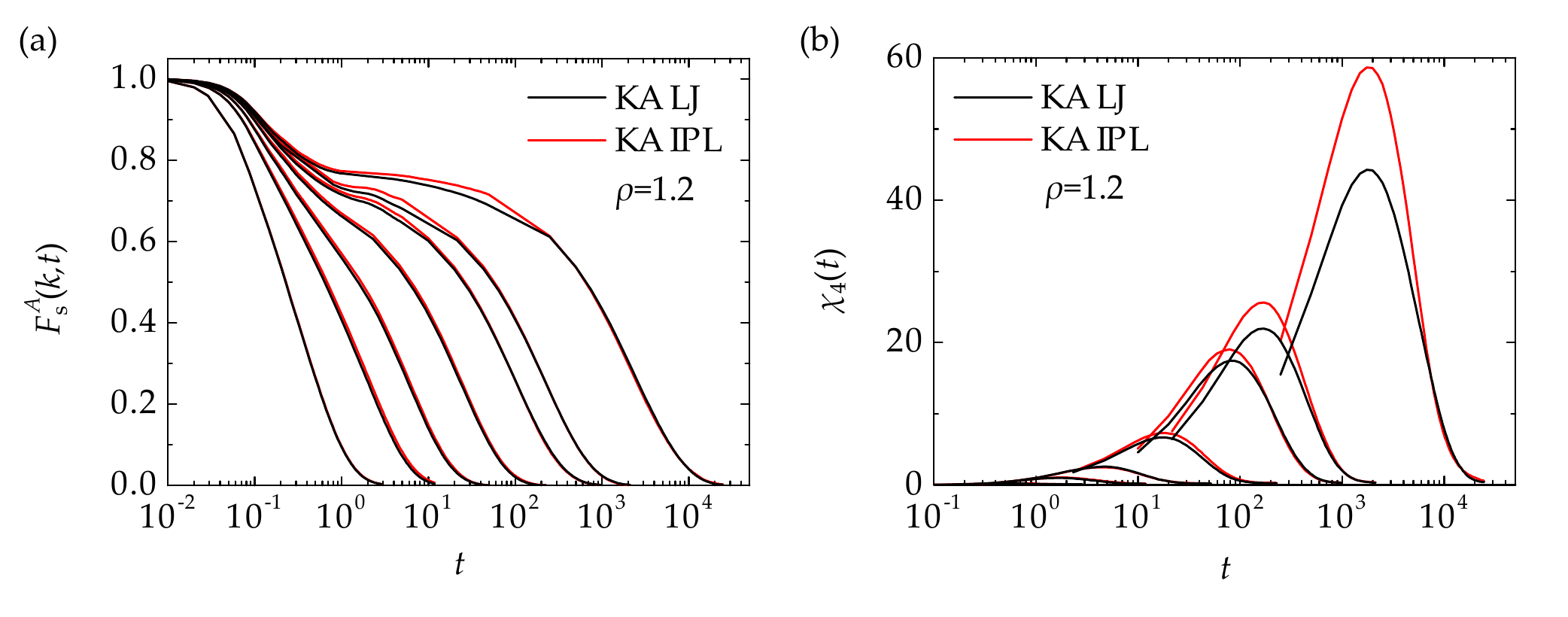}
\par\end{centering}
\caption{The dynamics of the KA LJ and IPL variants for $\rho=1.2$ and $T=\{2.0,1.0,0.75,0.6,0.525,0.5,0.45\}$}. The (a) self ISF $F_\mathrm{s}^A(k,t)$ and (b) dynamic susceptibility $\chi_4$. 
\label{figGdyn}
\end{figure*}

We proceed to consider the dynamics in the two systems by plotting the self intermediate scattering function for the $A$ particles $F_\mathrm{s}^A(k,t)$ and the dynamic susceptibility $\chi_4$ in Figs.~\ref{figGdyn}(a) and (b). The self ISFs are well matched between the two models, and consequently the fragility and degree of super-Arrhenius behavior in relaxation times are similar. This scenario contrasts with the WCA truncated approximation for the KA mixture, where the viscous behavior is found to be significantly different between the two models~\cite{berthier2009}. The difference between the relaxation dynamics of the KA LJ and KA-WCA models has been attributed to the numbers and spatial extent of 11A polyhedra \footnote{\noindent The TCC 11A cluster is equivalent to the Voronoi face (0,2,8) cluster identified in reference~\cite{coslovich2011}.} in the supercooled regime~\cite{coslovich2011}. 

We calculate the dynamic susceptibility $\chi_4(t)$ following La\v{c}evi\'{c} \emph{et al.~}\cite{lacevic2003}
\begin{equation}
\chi_4(t)=\frac{V}{N^2 k_B T}[\langle Q(t)^2\rangle -\langle Q(t)\rangle^2],
\label{eqChi4}
\end{equation}
\noindent where 
\begin{equation}
Q(t)=\frac{1}{N} \sum_{j=1}^N \sum_{l=1}^N w(|\textbf{r}_j(t+t_0)-\textbf{r}_l(t_0)|).
\label{eqQ}
\end{equation}
\noindent The overlap function $w(|\textbf{r}_j(t+t_0)-\textbf{r}_l(t_0)|)$ is defined to be unity if $|\textbf{r}_j(t+t_0)-\textbf{r}_l(t_0)|\le a$, 0 otherwise, where $a=0.3$. 

In Fig.~\ref{figGdyn}(b) we see that, except for the lowest temperatures, the dynamic susceptibilities are also well matched, indicating that the degree of cooperativity in the relaxation dynamics are comparable between the models. At the lowest temperatures, a discrepancy between the IPL and Lennard-Jones models appears which grows upon cooling. While finite size effects cannot be ruled out for these system sizes~\cite{karmakar2009}, the difference between the two systems suggests that the nature of the dynamic heterogeneity may be somewhat different between the two systems. In particular, it indicates that the number of particles in a given dynamically ``slow'' region may be higher in the case of the IPL.

\begin{figure*}
\begin{centering}
\includegraphics[width=18 cm]{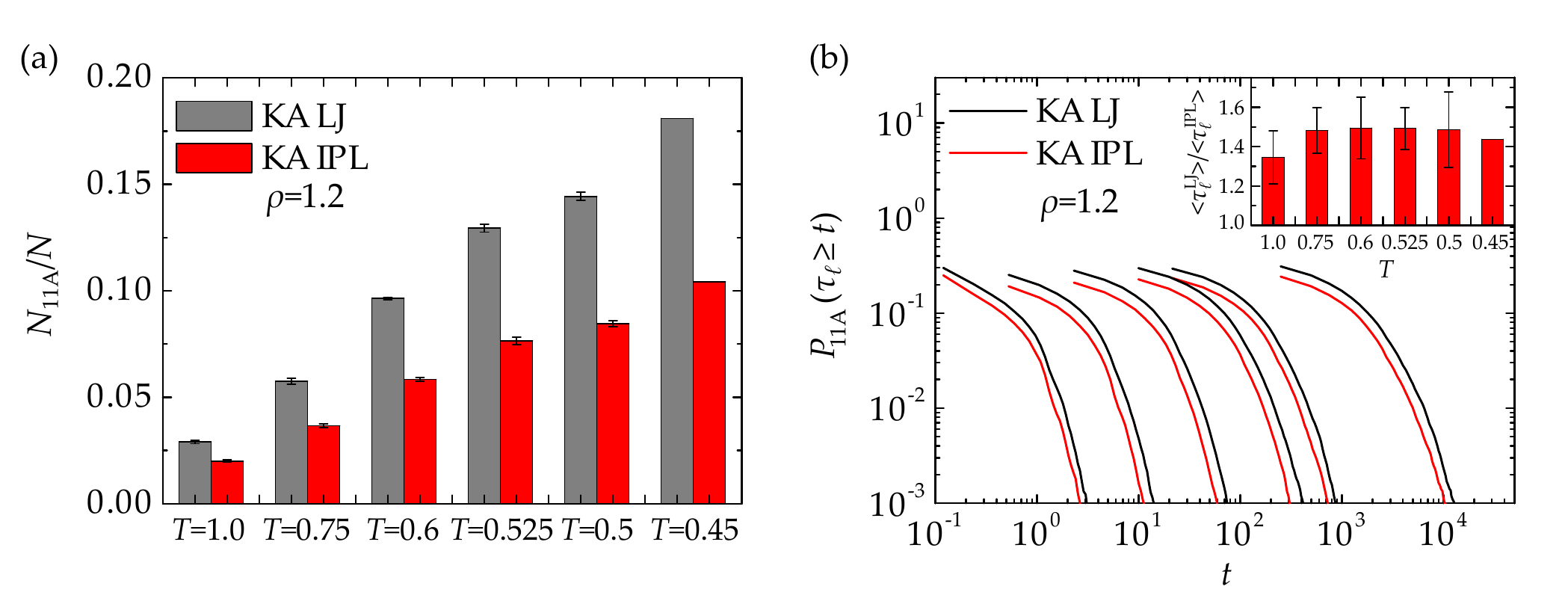}
\par\end{centering}
 \caption{Analysis of the statistics and dynamics of 11A clusters in the 
KA LJ and KA IPL mixtures for $\rho=1.2$ and $T=\{1.0,0.75,0.6,0.525,0.5,0.45\}$. (a) 11A population $N_\mathrm{11A}/N$ and 
(b) $P_\mathrm{11A}(\tau_\ell \ge t)$. Inset shows the ratio of the lifetimes of 11A clusters in the LJ and IPL systems,  $\langle\tau_\ell^\mathrm{LJ}\rangle/\langle\tau_\ell^\mathrm{IPL}\rangle$. Error bars on the bar charts show plus and minus two standard deviations from the mean.
\label{figGCompare}}
\end{figure*}

So far, measures like the pair correlation function have proven insufficiently sensitive to resolve the nature of the subtle differences in structure that exist between a Lennard-Jones liquid and its power-law counterpart, and which increase with cooling. However, it is clear from Fig.~\ref{figGCompare}(a), that the fraction of 
particles found in 11A clusters can differ dramatically, the fraction being almost twice as large for the Lennard-Jones liquid as compared to the model system. This difference increases strongly with cooling, and the subtle changes in structure become more pronounced as the system enters the glassy regime. As a measure of dynamical properties, we also looked at the {\it lifetime} of 11A clusters in Fig.~\ref{figGCompare}(b). Again, there is a discrepancy between the two models, with the 11A in the KA LJ system displaying lifetime autocorrelation functions that decay more slowly than the KA IPL model. The ratio $\langle\tau_\ell^\mathrm{LJ}\rangle/\langle\tau_\ell^\mathrm{IPL}\rangle$ gives a measure of the ratio of the average lifetime of 11A in the LJ system with respect to the IPL reference system [inset of Fig.~\ref{figGCompare}(b)]. The ratios indicate that the 11A lifetimes in the LJ are around 40\% longer than for the IPL system.

The qualitative differences in numbers of particles participating in 11A clusters between the two models may already be seen in subtle deviations in the partial radial distribution functions (Fig.~\ref{figG}). The $AB$ partial radial distribution function has a more pronounced first peak for the KA LJ mixture than the IPL reference system [Fig.~\ref{figG}(b)]. Conversely, the $B$ particles in the KA LJ model appear to have a lower affinity for bonds with other $B$-particles compared to the IPL counter part [lower first peak in Fig.~\ref{figG}(c)]. Both these deviations are consistent with a larger number of 11A clusters in the KA LJ model, as 11A clusters contain a central $B$-particle surrounded by large numbers of $A$-species and is a stable local arrangement of the particles~\cite{malins2013fara}. Once more, this shows that pair correlation functions are 
too blunt a tool resolve potentially important signatures of local structure and dynamics, whose correct interpretation is possible only once it has been picked up through an analysis of locally preferred 
structures.

\section{Discussion}
\label{sectionDiscussion}

These results raise a number of questions about the relationship between local structuring and glassy behavior. The bicapped square antiprism 11A cluster has been correlated with the non-Arrhenius dynamical behavior in the KA system~\cite{coslovich2007,malins2013fara}. The non-perturbative effect of attractions on the viscous behavior between the KA LJ and KA-WCA model was previously rationalized by differences in higher-order structural correlations between the two systems~\cite{coslovich2011}. However here we have shown that the repulsive IPL system, which matches the static and dynamic pair-correlations of the KA LJ system, displays markedly different behavior in terms of the static numbers and persistence of 11A clusters. The 11A cluster was suggested to be a proxy for higher-order structural correlations that may be necessary for inclusion in theories of the glass transition in order that the predictions of the relaxation times are accurate for both the  KA LJ and KA-WCA models~\cite{coslovich2011,berthier2011f}. 11A clusters have
 furthermore been identified with dynamically slow regions~\cite{malins2013fara}, a structural-dynamical phase transition in the KA system~\cite{speck2012}, and in ultra-stable glasses prepared vapor deposition~\cite{singh2013}. 

Here we find two similar models --  KA LJ and KA IPL -- where the composition, interactions, pair-structure, fragility and relaxation times are quantitatively similar, yet the 11A population (the proxy for higher-order structural correlations) shows considerable deviations. This emphasizes that low-order correlations do not fully describe glassy behavior.  Here, although two-point temporal and spatial correlation functions are very similar between the two systems, the four-point dynamic susceptibility $\chi_4(t)$ does show a tantalizing distinction at the lowest temperatures studied [Fig.~\ref{figGdyn}(b)]. In the population and lifetime of 11A clusters we propose a simple statistical measure that permits detection of \emph{qualitative} differences between the two models. Not only is the higher-order structure different between these two models [Fig.~\ref{figGCompare}(a)] but we have revealed a discrepancy in the 11A lifetime at all temperatures in Fig.~\ref{figGCompare}(b). 
This shows that local relaxation behavior is different between the two models.

We offer the following explanations for the discrepancies we find. Given that other glassformers, such as the Wahnstr\"{o}m binary Lennard-Jones model~\cite{malins2013jcp} and hard spheres~\cite{royallAngell} have been shown to exhibit other  locally favored structures, it is possible that the locally favored structure of the IPL might be another type of cluster with similar prevalence and lifetime to the 11A clusters in the KA LJ mixture. Indeed, the discrepancy in the dynamic susceptibility $\chi_4(t)$ [Fig.~\ref{figGdyn}(b)] could be interpreted that, if more particles are involved in dynamically slow regions in the KA IPL mixture, then a structural motif other than 11A should play a role. Now the topology of the ground-state cluster is strongly sensitive to interaction range~\cite{doye1995}. Here the power of the IPL potential ($15.48$), is rather higher that the 6-12 of the Lennard-Jones potential, so a change in locally favored structure is entirely possible -- even in the case that the pair structure \emph{and dynamics} are similar. Previously, we found that equilibrium liquids with interactions of differing range, mapped to one another, showed a strong change in the population of TCC clusters~\cite{taffs2010jcp}. If indeed the IPL has another locally favored structure, it would indicate that both two-point structure matching as the WCA approach~\cite{weeks1971,berthier2011f} and two-point matching of structure and dynamics in the case of strongly-correlating liquids~\cite{pedersen2010c} is insufficient to resolve detailed structure.

Another possibility to explain the differences we observe in the 11A populations is that the mapping between the IPL and full KA LJ may break down at lower temperatures, and that we are starting to see signs of this for our lower temperatures. Ultimately breakdown is unavoidable, as the correlation coefficient for $U-W$ fluctuations decreases linearly with $1/T$ to a good approximation [Fig.~\ref{figGRIsomorphs}(a)] and eventually KA LJ will no longer be strongly correlating~\cite{dyre2013a}.

Although we find differences in higher-order correlations of structure and dynamics in terms of 11A clusters between the full Lennard-Jones system and the IPL reference model, others have found better agreement using different measures for higher-order correlations. In particular, Hocky \textit{et al.\ }measured ``point-to-set'' correlation lengths down to $T=0.55$ and found good agreement between the two systems~\cite{hocky2012}. Recently a different length scale extracted from the cross-over where the lowest eigenvalue of the Hessian matrix has been shown to be sensitive to disorder, is identical to the point-to-set length up to a rescaling factor~\cite{karmakar2012,biroli2013}. The second length scale is more readily accessible at lower temperatures than the point-to-set length, and the agreement between the two systems holds down to $T=0.43$. Unlike the TCC we use, these approaches are ``order-agnostic'', and insensitive to a particular structure. Thus an identical structural lengthscale in these systems is entirely consistent with our suggestion that another structural motif apart from the 11A bicapped square antiprism might play a role in the KA IPL mixture, given that relative scarcity of 11A. Moreover, we have found the same static length in the KA LJ mixture using the TCC ~\cite{malins2013fara} as that found using the order-agnostic approaches ~\cite{hocky2012,karmakar2012,biroli2013}. The differences we have found underline the ability of the TCC to identify particular structural motifs. Thus comparison of the Lennard-Jones and IPL systems presents the opportunity to study the strengths and weaknesses of various higher-order correlations that are proposed to be of importance for glassy dynamics.

\section{Conclusions}
\label{sectionConclusion}

Here we have considered the behavior of structural correlations in a strongly correlating liquid and compared them to its perfectly correlating inverse power law reference system. We demonstrated with measurement of the statistics and dynamics of bicapped square antiprism 11A clusters along isomorphs of the Kob-Andersen Lennard-Jones mixture that these higher-order structural and dynamical correlations show very much larger deviations along the Lennard-Jones isomorphs than do two-body correlations. This result is at odds with the invariance of structure in reduced units predicted by the theory of isomorphs. The size of the deviations in the higher-order structural correlation functions between isomorphic state points increases as the coefficient of correlation $R$ decreases on cooling.

The temperature behavior of 11A clusters in the KA LJ model was compared with its IPL reference potential. Deviations of up to 80\% were found in both the fraction of particles participating in 11A clusters and the lifetime of 11A clusters, indicating that differences in the higher-order structural and dynamic correlations do not necessarily translate to differences in relaxation times. It remains an open question as to the importance of 11A clusters in determining the glassy behavior of the KA IPL model. In particular, we suggest that because of the sensitivity of higher-order structure to interaction range, other local structures may play a role in the IPL reference system, given that it has a shorter range than the LJ model. To settle some the issues raised it would be necessary to perform a more detailed study of the behavior of higher-order structural correlation functions in the KA IPL mixture. It would also be beneficial to consider the lower temperature behavior of the KA LJ model and its IPL counterpart to understand how the viscous dynamics develop in each case.

\subsection*{Acknowledgements}
We thank Mark Pond and Stephen Williams for stimulating discussions at the start of this project. Bob Evans is acknowledged for his enlightening opinions on isomorphs. Jeppe Dyre is thanked for helpful comments on this work. A.~M.\ is funded by EPSRC grant code EP/E501214/1. C.~P.~R.\ thanks the Royal Society for funding. 
This work was carried out using the computational facilities of the Advanced Computing Research Centre, University of Bristol.


%

\end{document}